\documentstyle[11pt]{article}

\title{The Information Manifold for Relatively Bounded Potentials}
\author{R. F. Streater\\Dept. of Mathematics, King's College London,\\
Strand,London WC2R 2LS}
\date{2/8/1998}
\newtheorem{theorem}{\bf Theorem}
\newtheorem{definition}[theorem]{\bf Definition}
\newtheorem{lemma}[theorem]{\bf Lemma}

\newtheorem{corollary}[theorem]{\bf Corollary}
\begin{document}
\maketitle
\begin{abstract}
We construct a Banach manifold of states, which are Gibbs states for
potentials that are form-bounded relative to the free Hamiltonian. We
construct the $(+1)$-affine structure and the $(+1)-$connection.\\
{\bf keywords} Information manifold, Fisher metric, quantum geometry,
Bogoliubov metric, Kubo theory, statistical manifold.
\end{abstract}
\section{Introduction}

The ultimate goal of the present project is
a quantum version of the theory of information (or statistical) manifolds;
in classical probability this circle of ideas has been rather successful
in many fields from estimation theory to
dissipative dynamics in neural networks \cite{Amari2}. We were inspired
by the nice work of Pistone and Sempi, who put the classical theory
on a firm mathematical foundation in the infinite-dimensional case
\cite{Pistone}. While our ambitions are the same as \cite{Pistone},
our results are not as complete, and the technical problems, arising
from the non-commutative nature of the potentials, are quite different.
The problem first arose in the quantum theory of many-body systems
at `finite' temperature, in the works of Matsubara, Mori and Kubo
\cite{Matsubara,Mori,Kubo}. There, we find the correlation functions
for observables written as imaginary-time-ordered products of operators.
The mathematical theory was advanced by Bogoliubov, \cite{Bog} who showed
that the two-point correlation function was
real and positive-definite. In the geometrical context, this is
thus a Riemannian metric on the vector space of Hermitian operators;
it is often known as the Bogoliubov-Kubo-Mori, or {\em BKM}, metric.

Parametric estimation theory starts with a family of probability
distributions, and also the data which give us a distribution in the form
of a histogram. We seek the best representation the data; this
is done by finding the member of the family that minimises the `distance'
to the distribution of the data; here, which concept of distance to be used
is one of the problems. Gauss
used the Euclidean distance, giving his famous least-squares fit to the
data. This distance, however, depends on the coordinates used. Fisher
\cite{Fisher} introduced the {\em information matrix}, which is a tensor
under change of coordinates, and this was developed into a Riemannian metric
tensor on the manifold of parametrised distributions by Rao \cite{Rao}.
Dawid \cite{Dawid} realised that the theory also needed an affine
connection, and that this did not have to be that of Levi-Civita. Ideas from
information theory were then incorporated, and the `distance' to be
minimised turned out to be the Kullback-Leibler relative entropy.
The poetic geometry involving the dual affine structures was then
put together by Amari, in a notable book \cite{Amari1}. The manifold of
states is determined by a chosen subspace ${\cal X}$, spanned by
(linearly independent) random variables $\{X_1,\ldots,X_n\}$. It can be
parametrised by the canonical coordinates $\xi^j$, $\xi\in V$,
where $V$ is a convex cone in ${\bf R}^n$.
Let $X=\sum\xi^jX_j$; the {\em exponential family} determined by ${\cal X}$
is the set of states of the form 
\begin{equation}
p_{_X}=Z^{-1}\exp\{-\sum\xi^jX_j\}=\exp-\{X+\Psi(\xi)\}.
\end{equation}
The canonical coordinates are (inverses of) generalised temperatures
\cite{Ingarden2} and the Massieu function $\Psi=\log Z$
is the thermodynamic potential. 
The $(+1)$-affine structure comes from forming convex mixtures of the
$\xi^j$; that is, the mixtures of $p_{_X}$ and $p_{_Y}$ are states of the
form $p_{_{\lambda X+\lambda^\prime Y}}$. In this paper, $\lambda^\prime$
will denote $1-\lambda\geq 0$.
Since the $\xi$ are global affine coordinates, this affine structure is flat
and torsion-free. The Legendre dual to $\Psi$ is the entropy, written
as a function of the probability, and it also plays the role of a
thermodynamic potential, The dual coordinates,
\begin{equation}
\eta_i= -\frac{\partial\Psi}{\partial\xi^i}
\end{equation}
are the expectation values of the random variables $X_i$ in the state
$p_{_X}$.

The $\eta_i$ are also global coordinates, and the $(-1)$-affine structure
is defined by forming convex sums of these coordinates; it is therefore
also flat and torsion-free. It coincides with the usual convex mixture
of states, and so is called the {\em mixture} affine structure.
Each of these affine structures defines a
concept of parallel transport of vectors in the tangent space; neither of
these affine structures is
metric invariant, but they are dually related by the metric.
Amari also interpolates between these, to get the family of
$(\alpha)$-affine structures, of which $\alpha=0$ is self-dual and therefore
metric: it is the Levi-Civita affine structure.

It has been remarked \cite{Jaynes,Kossakowski,Ingarden,Balian,Amari2,RFS3}
that several dissipative models used in neural nets and
physics can be expressed as the projection or rolling of a linear
dynamics onto
the surface given by a family of distributions. The random variables
whose distributions
form the manifold are taken to be the slow variables of the theory;
the other variables are thermalised by the projections that keep the
dynamics on the manifold.

The mathematical legitimacy of the procedure was strengthened by the studies
of Chentsov \cite{Chentsov}. He had the idea of regarding stochastic maps as
the natural
morphisms between statistical structures. The Fisher-Rao metric is
contracted by any such map; moreover, it is the only Riemannian metric
(up to a constant factor) to have this property. In physics, dissipative
dynamics is given by a semigroup of stochastic maps, and the contractive
property is the expression of convergence to equilibrium at large times.
These are necessary properties of any good theory. Thus there is a certain
uniqueness, in the classical case, of the geometry of parametric families.
Chentsov remarked that this is not the case in the geometry of quantum
information manifolds
in $n$ dimensions. This was studied by Hasagawa, Nagaoka and Petz
\cite{Hasagawa1,Hasagawa2,Hasagawa3,Hasagawa4,Hasagawa5,Nagaoka,Petz1}.
See also \cite{Ohya}. The set of faithful density matrices is a
manifold ${\cal M}$ of dimension $n^2-1$. The morphisms are taken to be
stochastic completely positive maps. Chentsov showed that there are many
metrics on the tangent space ${\cal T}$ of ${\cal M}$ that are contracted by
these morphisms. One can identify ${\cal T}$ at $\rho$
with the linear space of Hermitian matrices with zero expectation in
the state $\rho$. This is the quantum analogue of the `centred random
variables' that make up the tangent space in $+1$ coordinates in the
classical case. Chentsov's work on the possible metrics was completed by
Petz \cite{Petz2} in the case of finite dimensions. Examples of these can be
found in Roepstorff \cite{Roepstorff}. Hasagawa, and Nagaoka, in particular
emphasise two important cases.
These are the symmetric {\em KMS} metric, and the {\em BKM} metric.

Given a faithful
density matrix, $\rho$, the {\em KMS} metric on the vector space of
$n\times n$ Hermitian matrices is constructed from the complex scalar
product $\langle X,Y\rangle={\rm Tr}\,(\rho X^*Y)$ by taking the real part.
The {\em KMS} metric has been used
in quantum estimation theory \cite{Uhlmann,Caves}, and it
coincides with the Study-Fubini metric on the projective sphere when
restricted to the pure states. I have also used it extensively in
\cite{RFS2}. In spite of this, it seems that the {\em BKM} metric is
better. It is defined on the space of {\em centred} operators $\widehat{X}=
X-\rho.X$ by
\begin{equation}
\langle\widehat{Y},\widehat{X}\rangle:=\mbox{Tr}\int_0^1
\rho^{\lambda}\widehat{Y}\rho^{\lambda^\prime}\widehat{X}d\lambda
\end{equation}
Here, and elsewhere in the paper, $\lambda^\prime=1-\lambda$.
First, in the quantum case, the $(\pm)$-affine structures
are not dual relative to the
{\em KMS} metric. Alternatively, one can take, as in \cite{Nagaoka}, the
mixture affine structure as a start; then its dual relative to the
{\em KMS} metric is not flat and torsion free. It follows that there
do not exist dual potentials, related by a Legendre transform,
corresponding to the physically important objects, the entropy and the
Massieu function. The second reason why I now prefer the {\em BKM} metric is
mathematical; the {\em BKM} metric is smaller than the {\em KMS} metric;
the latter does not exist in general for unbounded operators, and certainly
not for forms. The mathematical {\em stylishness} of the {\em BKM}
version of the information manifold is so compelling that perhaps the
extensive work on quantum estimation should be redone with the {\em BKM}
metric replacing the {\em KMS} metric.

In the classical case, Pistone and Sempi \cite{Pistone,Gibilisco} introduce
information manifolds
in general, not parametrised by a finite number of slow variables.
Thus they are in the field of {\em nonparametric estimation theory}.
However, they must start somewhere, and they fix a basic underlying measure
space, whose measure $\mu$ need not be finite, but is used to specify the
null sets. The probability measures of the manifold are then those smoothly
related to the
given one. The present paper is an attempt to get a nonparametric
version in the quantum case. It follows up earlier work \cite{RFS1}, in
which the Hilbert space was of infinite dimension, but
the manifold was of finite dimension, corresponding to limiting our
attention to finitely many `slow' variables.
In the quantum case, we need a trace, not necessarily finite, to play the
role of $\mu$; we need a density matrix to play the role of $p$;
this is provided by
a `free Hamiltonian' $H_0$, a positive selfadjoint operator
with domain ${\cal D}(H_0)$, on a Hilbert space ${\cal H}$, such that
there exist $\beta_0>0$ with
\begin{equation}
\rho_0=Z_0^{-1}e^{-\beta H_0} \mbox{ is a density operator for all }\beta>
\beta_0.
\label{stable}
\end{equation}

This condition holds for the harmonic oscillator, and also for the
Laplacian in a compact region in ${\bf R}^n$, with smooth boundary and
Dirichlet boundary conditions, or in a rectangle in ${\bf R}^n$ with
periodic boundary
conditions. In all these examples, $\beta_0=0$. The condition corresponds
to a thermodynamically stable system in a finite box, in which there are
no phase transitions for $\beta>\beta_0$. The zero-point energy of $H_0$
has no significance, as the
addition of a constant to $H_0$ is cancelled by the corresponding change in
$Z$; we may therefore assume that $H_0\geq I$. By scaling, we may assume
that $\beta_0<1$, and we start with a state of the form
\begin{equation}
\rho_0=e^{-(H_0+\Psi_0)}
\end{equation}
Here, $\Psi_0=\log Z_0$. The condition given by eq.~(\ref{stable}) is
enough to allow the quantum analogue of the `Cramer class' of random
variables $u$ arising in the classical case \cite{Pistone}. 

In \S(2) we shall construct our first patch of the manifold.
It will be a set of states related to the basic state $\rho_0$
by a small form perturbation of $H_0$. A {\em form}
is a bilinear real map $\varphi,\varphi\mapsto X(\varphi,\varphi)\in{\bf R}$,
where $\varphi$ runs over the form-domain $Q(X)$. For example, the positive
selfadjoint operator $H_0$ defines the quadratic form $q_0$ with form domain
$Q_0=D(H_0^{1/2})$ by the definition
\begin{equation}
q_0(\varphi,\varphi)=\langle H_0^{1/2}\varphi,H_0^{1/2}\varphi\rangle.
\end{equation}
The theory of small forms
allows us to write the operator $H_X$ as the unique selfadjoint
operator satisfying
\begin{equation}
\langle H_X^{1/2}\varphi,H_X^{1/2}\varphi\rangle=q_0(\varphi,\varphi)
+X(\varphi,\varphi),
\end{equation}
for all $\varphi\in Q_0$. The perturbed state
\begin{equation}
\rho_X=Z_X^{-1}\exp-\{H_X\}=\exp-\{H_X+\Psi_X\}
\label{states}
\end{equation}
is shown (lemma 4) to exist provided that the {\em form bound} of $X$ is
smaller than $1-\beta_0$; it inherits most of the good properties of
$\rho_0$. The forms $X$ which are $q_0$-bounded give us the Cramer class at
$\rho_0$.

States of the form $\rho_X$ are the canonical states for
the Hamiltonian $H_0+X$. The case of bounded perturbations has been
extensively analysed by Araki \cite{Araki}. In linear response theory
such states are thought
of as the equilibrium state reached in response to an external field,
whose `effective potential' is $X$. We do not insist on this
interpretation; for example, in the version of non-equilibrium statistical
mechanics called `statistical dynamics' \cite{RFS2} we regard $\rho_X$
as a nonequilibrium state parametrised by $X$. Ingarden \cite{Ingarden}
has called the possible $X$ the `slow variables'; Jaynes calls them
the accessible variables, in line with his subjective view of entropy.
The point of working in infinite dimensions is to have a space of states
large enough to contain the dynamics, so that the `reduced description'
can be given a geometric flavour as the projection from the full manifold
to a submanifold described by the means of the variables of interest.

The parametrisation of the perturbed states is established by an
excursion into the theory of {\em sesqiforms}. We show that if $X$
is relatively form-bounded, then the expectation value
$\rho_0.X={\rm Tr}\,(\rho_0X)$ can be given an unambiguous meaning, and
is continuous in $X$ when the space of relatively bounded forms is
provided with with a natural norm, in which it becomes a Banach space,
${\cal T}(0)$. A form obeying $\rho_0.X=0$ is said to be {\em centred}.
The subspace $\widehat{\cal T}(0)\subseteq{\cal T}(0)$ given by centred
variables then defines a closed subspace; the open ball of radius
$1-\beta_0$ in $\widehat{\cal T}(0)$ is in bijection
with a collection of states of the form eq.~(\ref{states}). This is our
first {\em patch} of the information manifold, i.e. mapping from the set
of states into the open unit ball of a Banach space.

In \S(3) we develop analysis with the first patch; in particular, we derive 
the Duhamel formula for the difference of two states of the
form eq.~(\ref{states}) in terms of integrals of sesqiforms.

In \S(4) we look at the two affine structures $(\pm1)$ in the first
patch. Parallel transport in the $(+1)$ structure is easy to define; but
the $(-1)$-affine sum, which is ordinary mixture of states, may lead
outside the manifold.

In \S(5) the manifold is extended by adding overlapping patches based
on points in already established patches. The key here is that a
perturbed state $\rho_X$ inherits all the good properties of $\rho_0$,
and that the
norms on overlapping patches are equivalent.
That we should be able to do better than just the first patch
is easy to understand; if $X=(1/2)H_0$, then $X$ is $H_0$-small, and we can
define $H_X=H_0+X=(3/2)H_0$. Then an operator $Y$ can be $H_X$-small without
$X+Y$ being $H_0$-small. So we define $H_0+X+Y=(H_0+X)+Y$ in two stages.
In this case we get the same operator for $\beta(H_0+X+Y)$ as if we use
$3/2\beta(H_0+2/3Y)$, so in this case we could get there in one step from a
state of different temperature. However, in general we expect to reach new
states, not obtainable in one step. In this way we build up our manifold
of states, reachable from $\rho_0$ in a finite number of steps. It is
clear that the whole space of $H_0$-bounded operators cannot be reached
from the state of given beta; for $X=-H_0$ will never be reached; roughly,
the manifold we construct lies in the direction of $+H_0$.
The $(+1)$
affine structure and parallel transport can be extended to the whole
manifold, which is proved to be convex.

\section{Sesquiforms and Perturbation Theory}

In this section we extend the definition of perturbed states, beyond
that considered in \cite{RFS1}, in two ways.
First, we allow $X$ to be a quadratic form, bounded relative to the quadratic
form $q_0$.
This means the following. Let $X$ be a sesquilinear form defined on $Q_0$;
it is said to be $q_0$-bounded if there exist numbers, $a$, $b$ such that
\begin{equation}
|X(\psi,\psi)|\leq a q_0(\psi,\psi)+b\|\psi\|^2 \mbox{ for all }\psi\in Q_0.
\end{equation}
If $a$ can be chosen less than $1$ (by a good choice of $b$), then we
say that $X$ is $q_0$-small. In our case, we shall need to choose $a$
smaller than $1-\beta_0$. If $a$ can be chosen to be arbitrarily small,
we say that $X$ is $q_0$-tiny. 

It is not hard to show that $D(H_0)\subseteq Q_0$.
Any $H_0$-small operator is also $q_0$-small \cite{Reed}, Th X.18. 

The second extension of \cite{RFS1} is that we consider simultaneously
the set of all $q_0$-bounded forms, and provide them with a norm; 
we obtain a parametrisation of the space of perturbed states by $X$ of
small norm by the open unit ball of a Banach space: this is our first patch
of the manifold.
\subsection{Sesquiforms}
We can give a meaning to left and right products of quadratic forms by
certain operators. Suppose that $q$ is a quadratic form with domain
$Q(q)$; then $q$ defines a sesquilinear form $q(\phi,\psi)$ by
polarisation, with domain $Q(q)\times Q(q)$. Let $A,B$
be operators such that $A^*$ and $B$ are densely defined, $A^*$ taking
$D(A^*)$ into $Q(q)$, and $B$ mapping $D(B)$ into $Q(q)$.
Then by the expression $AqB$ we mean the sesquilinear form given by
\[\phi,\psi\mapsto q(A^*\phi,B\psi),\hspace{.4in}\phi\in D(A^*),
\hspace{.2in}\psi\in D(B).\]
It is obvious that the product is associative: $(AB)q=A(Bq)$.
More generally, given dense linear sets ${\cal D}_1$, ${\cal D}_2$,
a sesquilinear map from ${\cal D}_1\times{\cal D}_2$ to ${\bf C}$
will be called a {\em sesquiform}. A sesquiform is not required to be
symmetric. The `formal adjoint' of the sesquiform $q$ is the form $q^*$
with domain $D_2\times D_1$, and given by
\begin{equation}
q^*(\varphi,\psi)=\overline{q(\psi,\varphi)}.
\end{equation}
We note that the restriction of a sesquiform to a pair of dense linear
subspaces of ${\cal D}_1$ and ${\cal D}_2$ is also a sesquiform, and
that sesquiforms with the same domains can be added to give a
sesquiform.
\begin{definition}
A sesquiform $q$ will be said to be bounded if there exists $C$ such
that
\[|q(\varphi,\psi)|\leq C\|\varphi\|\,\|\psi\|\mbox{ holds on the domain.}\]
\end{definition}
\begin{lemma}
Suppose that $X$ is a $q_0$-bounded symmetric form defined on $Q_0$.
Then $R_0^{1/2}XR_0^{1/2}$ is a bounded symmetric form defined
everywhere. Conversely, if $X$ is a symmetric form with domain $Q_0$, and
$R_0^{1/2}XR_0^{1/2}<1$, then $X$ is $q_0$-small.\label{small}
\end{lemma}
PROOF.\\
Recall that we have normalised $H_0$ so that $H_0\geq I$; then $R_0=
H_0^{-1}$ is bounded (by 1). It is known that $R_0^{1/2}$ maps ${\cal H}$
onto $Q_0$, and so $R_0^{1/2}XR_0^{1/2}$ is everywhere defined. So, suppose
that $X$ is $q_0$-bounded. Then
\begin{eqnarray*}
\left|R_0^{1/2}XR_0^{1/2}(\psi,\psi)\right|&=&\left|X(R_0^{1/2}
\psi,R_0^{1/2}\psi)\right|\\
&\leq&aq_0(R_0^{1/2}\psi,R_0^{1/2}\psi)+b\|R_0^{1/2}\psi\|^2\\
&\leq&(a+b)\|\psi\|^2
\end{eqnarray*}
so $R_0^{1/2}XR_0^{1/2}$ is bounded.

For the converse, assume that $X$ is a quadratic form with domain $Q_0$,
and that $a=\|R_0^{1/2}XR_0^{1/2}\|<1$, and let $\psi=R_0^{1/2}\varphi$
be any element of $Q_0$. Then
\begin{eqnarray*}
X(\psi,\psi)&=&\langle\varphi,R_0^{1/2}XR_0^{1/2}\varphi\rangle\leq
a\|\varphi\|^2\\
&=&aq_0(\psi,\psi).
\end{eqnarray*}
Hence $X$ is $q_0$-small, with $b=0$.\hspace{\fill}$\Box$

The Kato-Rellich theory can be extended to forms. The key is
the {\em KLMN} theorem, (\cite{Reed}, Vol. 2, p167)
which we give here in weaker form.
\begin{theorem}
Let $H_0$ be a positive self-adjoint operator, with quadratic form $q_0$
and form domain $Q_0$; let $X$ be a $q_0$-small symmetric quadratic form.
Then there exists a
unique self-adjoint operator $H_X$ with form domain $Q_0$ and such that
\begin{equation}
\langle H_X^{1/2}\varphi,H_X^{1/2}\psi\rangle=q_0(\varphi,\psi)+
X(\varphi,\psi),\hspace{.2in}\varphi,\psi\in Q_0.
\end{equation}
Moreover, $H_X$ is bounded below by $-b$, and any domain of essential
self-adjointness for $H_0$ is a form core for $H_X$.
\end{theorem}

Now let $X$ be $q_0$-small, with bounds $a,b$, with $a<1-\beta_0$. Denote
by $H_X$ the unique operator given by $KLMN$; let $q_X$ denote its form.
$H_X$ inherits the main property of $H_0$, thus:
\begin{lemma}
$\exp(-\beta H_X)$ is of trace class for all $\beta>\beta_{_X}=
\beta_0/(1-a)$.
\end{lemma}
PROOF\\
We have, as quadratic forms on $Q_0$, the inequalities
\begin{equation}
-bI+(1-a)q_0\leq q_X\leq bI+(1+a)q_0.
\label{eigen}
\end{equation}
Let $L$ be any finite-dimensional subspace of $Q_0$, and let $q$ stand
for $q_0$ or $q_X$. Put
\begin{equation}
\lambda(q,L)=\sup\{q(\psi,\psi):\|\psi\|=1,\psi\in L\}.
\end{equation}
Then the ordered eigenvalues of $q$ are given by the Rayleigh-Ritz
principle:
\begin{equation}
\lambda(q,n)=\inf\{\lambda(q,L):\dim L=n\}
\end{equation}
From the inequality (\ref{eigen}) we have for each subspace $L$
\begin{equation}
-b+(1-a)\lambda(q_0,L)\leq\lambda(q_X,L)\leq b+(1+a)\lambda(q_0,L).
\end{equation}
Taking now the inf over $L$ we get \cite{EBD}.
\begin{equation}
-b+(1-a)\lambda(q_0,n)\leq\lambda(q_X,n)\leq b+(1+a)\lambda(q_0,n).
\end{equation}
Since $\lambda(q_0,n)\rightarrow\infty$ as $n\rightarrow\infty$,
the spectrum of $H_X$ is purely discrete. We thus get for any $\beta>0$,
\[e^{b\beta}e^{-(1-a)\lambda(q_0,n)\beta}\geq e^{-\lambda(q_X,n)\beta}\geq 
e^{-b\beta}e^{-(1+a)\lambda(q_0,n)\beta}.\]
Taking the sum over $n$ gives the traces:
\[e^{b\beta}Tr\left(e^{-(1-a)H_0\beta}\right)\geq Tr\left(e^{-H_X\beta}
\right)\geq e^{-b\beta}Tr\left(e^{-(1+a)H_0\beta}\right).\]
So if $\exp\{-(1-a)\beta H_0\}$ is of trace-class for all $(1-a)\beta>
\beta_0$, then $\exp\{-H_X\beta\}$ is of trace class for all
$\beta>\beta_X=\beta_0/(1-a)$ .\hspace{\fill}$\Box$.

We define the {\em Cramer class} at $\rho_0$ to be the $q_0$-bounded
forms $X$; for then by Lemma~(4) there exists a neighbourhood $N$ of
zero such that for $\lambda\in N$, $\exp\{-(H_0+\lambda X\}$ is of
trace-class.

It follows from lemma~(\ref{small}) that if
$\|R_0^{1/2}XR_0^{1/2}\|<a_o=1-\beta_0$, then $H_X>0$; for we may take
$b=0$ in the {\em KLMN} theorem.

\subsection{The First Piece}
We now get the first piece of our manifold.
Let ${\cal T}(0)$ be the real linear space of $q_0$-bounded quadratic forms,
with domain $Q_0$ and norm
\begin{equation}
\|X\|_0=\|R_0^{1/2}XR_0^{1/2}\|<\infty.
\end{equation}
We note that $I$, the unit operator, lies in ${\cal T}(0)$. The map
$A\mapsto H_0^{1/2}AH_0^{1/2}$ from the set of all bounded Hermitian
operators ${\cal B}({\cal H})$ onto the set of symmetric
$q_0$-bounded sesquiforms
is an isometry; this shows that ${\cal T}(0)$ is isometric to ${\cal B}
({\cal H})$; in particular, ${\cal T}(0)$ is complete, and so is a Banach
space.
To each element $X$ of the interior Int$\,{\cal T}_{a_o}(0)$ of the ball
in ${\cal T}(0)$ of radius $a_o=1-\beta_0$, define the density matrix
\begin{equation}
\rho_X=Z^{-1}e^{-(H_0+X)}=e^{-H_0+X+\Psi_X}.
\end{equation}
The first piece of our manifold is the set ${\cal M}_0$ of such states.
We set up the patch by mapping ${\cal M}_0$ bijectively onto the
interior of a ball in a Banach space; our space ${\cal T}(0)$, with
its ball ${\cal T}_{a_o}(0)$,
will not do, for if we alter $X$ by adding a multiple of $I$, we do not
change the state; $\rho_X=\rho_{X+\alpha I}$, as the change in $X$ just
leads to an equal and opposite change in $Z_X$, which cancels. Conversely,
if $\rho_X=\rho_Y$ lie in ${\cal M}_0$, then $X-Y$ is a multiple
of $I$. For, as $\rho_X,\rho_Y$ are faithful states, we may take
logarithms: $\log\rho_X=\log\rho_Y$. Then
\[H_0+X+\Psi_X=H_0+Y+\Psi_Y\]
so
\[X-Y=(\Psi_Y-\Psi_X)I.\]
Furnish ${\cal T}$ with an equivalence relation
\begin{equation}
X\sim Y\hspace{1in}\mbox{if }X-Y=\alpha I \mbox{ for some }\alpha\in{\bf R}.
\end{equation}
Then the equivalence classes are lines in ${\cal T}(0)$ parallel to $I$.
We furnish the set ${\cal T}(0)/\sim$ of equivalence
classes with the
topology induced from ${\cal T}$. That is, an open set in the quotient
is the set of all lines passing through some open set in ${\cal T}(0)$.
There is then a bijection between the
set ${\cal M}_0$ and the subset of this quotient
space defined by
\begin{equation}
\left\{\{X+\alpha I\}_{\alpha\in {\bf R}}:\|X\|_0<a_o=1-\beta_0\right\}.
\end{equation}
The bijection is given by
\begin{equation}
\rho_X\mapsto \tilde{\rho}_X=\{Y\in{\cal T}(0):Y=X+\alpha I\mbox
{ for some }\alpha\in{\bf R}\}.
\end{equation}
Thus ${\cal M}_0$ becomes topologised, by transfer of structure. It is
obviously a Hausdorff space. Indeed, it is well known that the quotient
topology is equivalent to the topology given by the quotient norm
\cite{Kato}, p140. However, to construct the patch, we parametrise
${\cal M}_0$ by a ball in a Banach subspace rather than the quotient.
In finite dimensions \cite{Hasagawa1,Nagaoka,Toth,Pistone}
this has been done by selecting a point on each line in $\widehat{\cal T}$,
namely, the centred variable $\widehat{X}=X-\rho_0.X$. 
The trouble is that we cannot prove that $\rho_0 X$ is a sesquiform
of trace class. We can however find a natural definition for its trace.
Suppose that $X\in{\cal T}(0)$, and consider the
sesquiform $\rho_0^\lambda X\rho_0^{\lambda^\prime}$ for $0<\lambda<1$.
Choose $\beta_1\in(\beta_0,1)$ and put $\delta_1=\lambda(1-\beta_1)$
and $\delta_2=\lambda^\prime(1-\beta_1)$. From associativity, this is
equal to
\begin{equation}
\left(\rho_0^{\lambda-\delta_1}\right)\left(\rho_0^{\delta_1}H_0^{1/2}\right)
\left(R_0^{1/2}XR_0^{1/2}\right)\left(H_0^{1/2}\rho_0^{\delta_2}\right)
\left(\rho_0^{\lambda^\prime-\delta_2}\right).
\label{bound1}
\end{equation}
This is an operator of trace class, as we see from the H\"{o}lder
inequality for Schatten norms:
\begin{equation}
\|ABCDE\|_1\leq \|A\|_{1/\lambda}\|B\|_\infty\|C\|_\infty\|D\|_\infty
\|E\|_{1/\lambda^\prime}
\end{equation}
where $A\ldots E$ are the factors bracketted in eq.~(\ref{bound1}).
For example, we note that
\[\|A\|_{1/\lambda}=\left(\mbox{Tr}\left|\rho_0^{\lambda-\delta_1}
\right|^{1/\lambda}\right)^\lambda=\|\rho_0^{\beta_1}\|_1^\lambda
<\infty\]
since $\beta_1>\beta_0$.
By cyclicity of the trace, its trace is
independent of $\lambda\in (0,1)$. This needs proving, because of the
possible existence of Connes cyclic cocycles. Formal differentiation of
$\rho_0^\lambda X\rho_0^{\lambda^\prime}$ gives
\[\rho_0^\lambda[X,H_0]\rho_0^{\lambda^\prime}\]
whose trace might be equal to
\[\mbox{Tr}\,\left(\rho_0XH_0-H_0\rho_0X\right)=0\]
by cyclicity; but it is just such expressions that might give non-zero
cyclic cocycles if not all the operators are bounded.

We hope to peel off a small
part of $\rho_0^{\lambda^\prime}$ and put it at the front. 
This would be possible if the remaining factor were of trace class.
This is proved in the following
\begin{lemma}
Suppose that $\rho_0^\beta$ is of trace class for all $\beta>\beta_0$, where
$0\leq\beta_0<1$; suppose that $X$ is $q_0$ bounded. Then
\[\rho_0^\lambda X\rho_0^{\lambda^\prime}\]
is of trace class for all $0<\lambda<1$, and its trace is independent of
$\lambda$; here as always, $\lambda^\prime=1-\lambda$.
\end{lemma}
PROOF.
The form is that of a bounded operator, since e.g. $\rho_0^
\lambda$ maps ${\cal H}$ into $Q_0$. Write
\begin{equation}
\rho_0^\lambda X\rho_0^{\lambda^\prime}=\rho_0^{\lambda-\delta_1}\left(
\rho_0^{\delta_1}H_0^{1/2}\right)\left(R_0^{1/2}XR_0^{1/2}\right)\left(
H_0^{1/2}\rho_0^{\delta_2}\right)\rho_0^{\lambda^\prime-\delta_2-
\delta}\rho_0^\delta
\label{Holder}
\end{equation}
for suitably chosen $\delta,\delta_1,\delta_2>0$, such that $\delta_1<
\lambda$ and $\delta+\delta_2<\lambda^\prime$.
The product of the three operators in brackets in eq.~(\ref{Holder}) is
bounded, by $C$ say (but not uniformly in the $\delta$'s). By the
H\"{o}lder inequality, this gives
\[\left\|\rho_0^\lambda X\rho_0^{\lambda^\prime-\delta}\right\|_1\leq
\left\|\rho_0^{\lambda-\delta_1}\right\|_{1/(\lambda-\delta_1)}.C.\left\|\rho_0
^{\lambda^\prime-\delta_2-\delta}\right\|_{1/\mu}\]
where $\lambda-\delta_1+\mu=1$, i.e. $\mu=\lambda^\prime+\delta_1$.
Now, $\rho_0$ is a positive operator, so
\[\|\rho_0^{\lambda-\delta_1}\|_{1/(\lambda-\delta_1)}=(\mbox{Tr}\,\rho_0)^
{\lambda-\delta_1}=1,\]
and
\[\|\rho_0^{\lambda^\prime-\delta_2-\delta}\|_{1/\mu}\]
is finite if
\[\rho_0^{(\lambda^\prime-\delta_2-\delta)/\mu}\]
is of trace class; the exponent is
\[(\lambda^\prime-\delta_2-\delta)/(\lambda^\prime+\delta_1)\]
which can be made larger than $\beta_0$ by choosing the $\delta$'s
very small. It follows that we can cycle the last factor, $\rho_0^\delta$,
to the front without changing the trace, thereby increasing $\lambda$
and decreasing $\lambda^\prime$. We can choose $\delta$ as close to
$\lambda^\prime-\delta_2$ as we please; it follows that the trace is
independent of $\lambda^\prime$ provided that $\lambda^\prime>\delta_2$.
Since $\delta_2$ was arbitrary, we have the result.\hspace{\fill}$\Box$

We define the {\em regularised mean} of $X$ in the
state $\rho_0$ to be
\begin{equation}
\rho_0.X:={\rm Tr}\left(\rho_0^{\lambda}X\rho_0^{\lambda^\prime}\right),
\mbox{ for one and hence all }\lambda\in(0,1).
\end{equation}
Moreover, $\rho_0.X$ is continuous as a map ${\cal T}(0)\rightarrow{\bf R}$.
This follows since our bound has $\|X\|_0$ as a factor.
The set
\begin{equation}
\widehat{\cal T}(0):=\{X\in{\cal T}(0):\rho_0.X=0\}
\end{equation}
is a closed linear subspace of ${\cal T}$ of codimension $1$. The norm
$\|\bullet\|_0$, restricted to $\widehat{\cal T}(0)$, makes $\widehat
{\cal T}(0)$ into a Banach space. We map ${\cal M}_0$ bijectively onto
the open subset of $\widehat{\cal T}(0)$ given by the points $\widehat{X}$
where the corresponding points in ${\cal T}(0)/\sim$ (i.e. the lines
$\tilde{\rho}_X$ parallel to $I$) intersect it; such a point is unique,
being given by $\alpha=-\rho_0.X$. This bijection is a homeomorphism, since
both ${\cal M}_0$ and $\widehat{\cal T}(0)$ have been given the topology
induced from ${\cal T}(0)$. The map $\rho_X\mapsto \widehat{X}$ is our first
chart and its inverse is our first patch.
As usual in the construction of Banach manifolds, we identify the tangent
space at the origin of this chart
with the space $\widehat{\cal T}(0)$ itself. In this identification,
the tangent of a curve of the form
\[\{\rho(\lambda)=e^{-\left(H_0+\lambda X+
\Psi_{\lambda X}\right)}:\lambda\in[-\delta,\delta]\}\]
is identified with $\widehat{X}=X-\rho_0.X$.
We see from the picture that our patch is the ``shadow'' of ${\cal T}_{1-
\beta_0}(0)$ onto the hyperplane $\widehat{\cal T}(0)$, and that it
contains the ball $\widehat{\cal T}_{1-\beta_0}(0)$ and in its turn is
contained in the open set
\[\{Y\in\widehat{\cal T}_1(0):\|Y\|_0<1+|\rho_0.Y|\}.\]

We note that in finite dimensions, the set of operators parallel to $I$
is orthogonal to the hyperplane $\widehat{\cal T}(0)$ when the space ${\cal
T}(0)$ is furnished with the {\em BKM} metric. We seem to need more
regularity than we have at present if the {\em BKM} metric is to be finite
in infinite dimensions. Obviously, $g_X(Y,I)$ can be defined when one of the
operators is the unit operator, and $g_X(Y,I)=\rho_X.Y$. Thus the
subspaces $\widehat{\cal T}(X)$ are all orthogonal to the space parallel
to $I$.
\section{Analysis in the First Patch}
So far we have a manifold ${\cal M}_0$ with one patch. Before enlarging the
manifold by the addition of more patches, we do some analysis.

First, it is clear that all states in ${\cal M}_0$ have finite entropy and
regularised mean energy, which are related by
\begin{equation}
S(\rho_{_X})=-{\rm Tr}\,\rho_{_X}\log\rho_{_X}=\rho_{_X}.H_{_X}+\Psi_X.
\end{equation}
For $\rho_X.H_X={\rm Tr}\left\{\rho_X^{1-\delta}\left(\rho_X^\delta H_X
\right)\right\}$ which is finite for $\delta<1-\beta_0$.

\begin{lemma}
Let $A$ be a closed operator and $B$ be a bounded operator such that
$B{\cal H}\subseteq D(A)$. Then $C=AB$ is bounded.
\label{easy}
\end{lemma}
PROOF.\\
We  note that $D(C)={\cal H}$, so by the closed graph theorem it is enough
to show that $C$ is closed. For this, let $\psi_n\rightarrow\psi$ and
suppose that $C\psi_n$ converges. Then we must show that $\psi\in D(C)$ and
$C\psi=\lim C\psi_n$. The first is true, as $D(C)={\cal H}$; for the second,
we see that $B\psi_n\rightarrow B\psi$, as $B$ is bounded, and $A(B\psi_n)$
converges. Since $A$ is closed, we conclude that $B\psi\in D(A)$
(already known) and $C\psi_n=A(B\psi_n)\rightarrow AB\psi=
C\psi$.\hspace{\fill}$\Box$

\begin{lemma}
Let $X\in{\cal M}_0$, $R_0=H_0^{-1}$ and $R_X=H_X^{-1}$. Then
$R_0^{1/2}H_X^{1/2}$ and $R_X^{1/2}H_0^{1/2}$ are bounded.
\label{easy2}
\end{lemma}
PROOF.\\
Since $H_0\geq I$, we see that $R_0^{1/2}$ is bounded and maps ${\cal H}$
into $D(H_0^{1/2})=Q_0=D(H_X^{1/2})$, and $H_X^{1/2}$ is closed. So by
lemma~(\ref{easy}), $C=H_X^{1/2}R_0^{1/2}$ is bounded; its adjoint,
the closure of $R_0^{1/2}H_X^{1/2}$, is therefore also bounded. We find
\[C^*C=R_0^{1/2}H_XR_0^{1/2}=1+R_0^{1/2}XR_0^{1/2},\]
so
\[\left(1-\|X\|_0\right)I\leq C^*C\leq\left(1+\|X\|_0\right)I.\]
Thus the inverse of $C$, namely $H_0^{1/2}R_X^{1/2}$, is
bounded by $\left(1-\|X\|_0\right)^{-1/2}$.\hspace{\fill}$\Box$

\begin{lemma}
Let $X$ and $Y$ lie in ${\cal M}_0$ and put $q_X=q_0+X$ on $Q_0$. Then
$Y$ is $q_X$-bounded.
\end{lemma}
PROOF.\\
As sesquiforms, we have
\begin{eqnarray*}
\left\|R_X^{1/2}YR_X^{1/2}\right\|&=&\left\|R_X^{1/2}H_0^{1/2}R_0^
{1/2}YR_0^{1/2}H_0^{1/2}R_X^{1/2}\right\|\\
&\leq&\left\|R_X^{1/2}H_0^{1/2}\right\|\;\|Y\|_0\;\left\|H_0^{1/2}R_X^
{1/2}\right\|\\
&<&\infty 
\end{eqnarray*}
by lemma~(\ref{easy2}).\hspace{\fill}$\Box$\\
We now come to the very useful Duhamel formula for forms.
\begin{theorem}
Let $X$ be a symmetric form, $q_0$-small, and
let $H_X$ be the
self-adjoint operator with form $q_0+X$. Then
\begin{equation}
e^{H_0}-e^{H_X}=\int_0^1e^{-\lambda H_0}Xe^{-\lambda^\prime H_X}d\lambda
\end{equation}
where the r.h.s. means the limit of $\int_\epsilon^{1-\delta}$
as $\epsilon$ and $\delta$ converge to zero of the given sesquiform
evaluated at any $\psi,\varphi\in{\cal H}\times{\cal H}$.
\end{theorem}
PROOF.\\
Consider the family of operators
\begin{equation}
F(\lambda)=e^{-\lambda H_0}e^{-(1-\lambda)H_X},
\end{equation}
where $0<\lambda<1$.
These are of trace-class, since we can apply H\"{o}lder's inequality
with parameters $1/\lambda$ and $1/\lambda^\prime$.
For any $\psi,\varphi\in{\cal H}$ we define
\begin{equation}
F_{\psi,\varphi}(\lambda)=\langle e^{-\lambda H_0}\psi,e^{-(1-
\lambda)H_X}\varphi\rangle.
\end{equation}
Since $e^{-\lambda H_0}$ maps ${\cal H}$ into $D(H_0)\subseteq Q_0$, and
$e^{-\lambda^\prime H_X}$ maps ${\cal H}$ into $D(H_X)\subseteq Q_0$, we see that
$F_{\psi,\varphi}$ is differentiable, and
\begin{eqnarray*}
\frac{d}{d\lambda}F_{\psi,\varphi}(\lambda)&=&-\langle H_0e^{-\lambda H_0}
\psi,e^{-(1-\lambda)H_X}\varphi\rangle+\langle e^{-\lambda H_0}\psi,H_X
e^{-(1-\lambda)H_X}\varphi\rangle\\
&=&-q_0\left(e^{-\lambda H_0}\psi,e^{-(1-\lambda)H_X}\varphi\right)
+q_X\left(e^{-\lambda H_0}\psi,e^{-(1-\lambda)H_X}\varphi\right)\\
&=&X\left(e^{-\lambda H_0}\psi,e^{-(1-\lambda)H_X}\varphi\right).
\end{eqnarray*}
Integrating from $0$ to $1$ gives the theorem.\hspace{\fill}$\Box$

\begin{lemma}
Suppose that $X,Y$ are $q_0$-bounded forms, and that the $q_0$-bound of
$Y$ is $a<a_o$. Then $\rho_0^\lambda X\rho_Y^{\lambda^\prime}$
is of trace class for $0<\lambda<1$.
\end{lemma}
PROOF: We can write 
\[\rho_0^\lambda X\rho_Y^{\lambda^\prime}=\rho_0^{\lambda\delta^\prime}
\left(\rho_0^{\lambda\delta}H_0^{1/2}\right)\left(R_0^{1/2}XR_0^{1/2}\right)
\left(H_0^{1/2}R_Y^{1/2}\right)\left(H_Y^{1/2}\rho_Y^{\lambda^\prime
\delta}\right)\rho_Y^{\lambda^\prime\delta^\prime}\]
where $\delta\in(0,1)$ will be chosen soon, and $\delta^\prime=1-\delta$.
The terms in brackets are all bounded, so the operator norm of their product
is bounded by $C$ say; this can grow as $\lambda$ approaches its limits
$0$ and $1$. We now use H\"{o}lder's inequality for traces, to get
\begin{eqnarray*}
\left\|\rho_0^\lambda X\rho_Y^{\lambda^\prime}\right\|_1&\leq&C\left\|
\rho_0^{\lambda\delta^
\prime}\right\|_{1/(\lambda\delta^\prime)}\left\|\rho_{_Y}^{\lambda^\prime\delta
^\prime}\right\|_{1/\mu}\\
(\mbox{where } \mu&=&\lambda^\prime+\delta\lambda)\\
&\leq&C\left\|\rho_0\right\|_1^{\lambda\delta^\prime}\,\left\|\rho_{_Y}
^{\lambda^\prime\delta^\prime/\mu}\right\|_1^\mu
\end{eqnarray*}
which is finite if
\[\left\|\rho_{_Y}^{\lambda^\prime(1-\delta)/(\lambda^\prime+\delta\lambda)}
\right\|_1<\infty.\]
Given $\lambda$ we can choose $\delta$ so small that
\[\frac{\lambda^\prime(1-\delta)}{\lambda^\prime+\delta\lambda}>\beta_{_Y}
=\beta_0/(1-a)\]
since the latter is less than 1.\hspace{\fill}$\Box$

This does not show that the integral converges in trace norm; for the
trace norm of the integrand might become unbounded at the ends, and
cannot be shown to be integrable over $[0,1]$. However, the trace, as
opposed to the trace-norm, does converge.
\begin{lemma}
Let $X,Y$ be $q_0$-small, $Y$ having  bound less than $a_0=1-\beta_0$. Then
${\rm Tr}\,\rho_0^\lambda X\rho_Y^{\lambda^\prime}$ is bounded for
$0<\lambda<1$.
\end{lemma}
PROOF.\\
We first note that we can use the cyclicity of the trace to take out
a factor $\rho_0^{\lambda\delta/2}$ on the left of the expression; this is
a bounded operator and the remaining product is, as above, still of
trace class. We can therefore permute these two factors under the trace.
We make use of this when $\lambda\geq 1/2$. If $\lambda<1/2$ we take out a
part of the power of the state $\rho_{_Y}$ from the right and put it on
the left under the trace; we now illustrate the method by doing this case.
\begin{eqnarray*}
\left|\mbox{Tr}\left(\rho_0^\lambda X\rho_{_Y}^{\lambda^\prime}\right)
\right|&=&\left|\mbox{Tr}\left\{\left(\rho_{_Y}^{\lambda^\prime\delta/2}
H_Y^{1/2}\right)\left(R_Y^{1/2}H_0^{1/2}\right)\rho_0^\lambda\right.\right.
\\
& &\left.\left.\left(R_0^{1/2}XR_0^{1/2}\right)\left(H_0^{1/2}R_Y^{1/2}
\right)\left(H_Y^{1/2}\rho_{_Y}^{\lambda^\prime\delta/2}\right)\rho_{_Y}
^{\lambda^\prime\delta^\prime}\right\}\right|.
\end{eqnarray*}
Since $\lambda^\prime\geq 1/2$, we have
\[\left\|\rho_{_Y}^{\lambda^\prime\delta/2}H_Y^{1/2}\right\|\leq
\sup_x\left\{x^{1/2}e^{-\delta x/4}\right\}\leq C /\delta^{1/2}\]
This bound occurs twice. The other factors in brackets are bounded operators
with norm bounded by $C_1$, independent of $\lambda$ and $\delta$. By
H\"{o}lder,
\begin{eqnarray*}
\left|\mbox{Tr}\left(\rho_0^\lambda X\rho_{_Y}^{\lambda^\prime}\right)\right|
&\leq&C_1C^2\delta^{-1}\left\|\rho_0^\lambda\right\|_{1/\lambda}\left\|
\rho_{_Y}^{\lambda^\prime\delta^\prime}\right\|_{\/\lambda^\prime}\\
&\leq&C_2\delta^{-1}.1.\left\|\rho_Y^{1-\delta}\right\|_1^{\lambda^\prime}.
\end{eqnarray*}
Now choose $1-\delta>\beta_{_Y}$ independent of $\lambda$. This gives a
bound on the trace independent of $\lambda\in(0,1/2)$. Similarly, in
the region $\lambda\in(1/2,1)$ we get a bound on the trace
independent of $\lambda$.\hspace{\fill}$\Box$
\begin{corollary} We have
\begin{equation}
{\rm Tr}\,e^{-H_0}-{\rm Tr}\,e^{-H_X}=\int_0^1{\rm Tr}\,
\left(e^{-\lambda H_0}Xe^{-\lambda^\prime H_X}\right)\,d\lambda.
\end{equation}
\end{corollary}
By inserting normalising factors we convert the exponentials into states,
and by specialising to the case $X=Y$ we show that the integrand
is a bounded function of $\lambda$ in $(0,1)$.
It follows that the integral
of the trace is absolutely convergent, and the trace is the sum of the
diagonal elements in any orthonormal basis. The trace and the $\int$ can be
exchanged, by Fubini's theorem.\hspace{\fill}$\Box$

We have obtained an estimate for the perturbation from the state $\rho_0$;
Now $H_X$ inherits the properties of $H_0$, at least if we replace it
by $H_X+I$. We have shown that if $Y$ obeys
\begin{equation}
\|Y\|_X:=\|R_X^{1/2}YR_X^{1/2}\|<a_{\rm x}
\end{equation}
then $Y\in{\cal M}_0$ is $q_X$-small; if $X$ is chosen small enough in
this norm, and depending on $X$, $Y$ is also small enough, we may replace
$H_0$ by $H_X$ and $H_X$ by $H_{X+Y}$ in these estimates. This may also
be done in lemma~(\ref{easy}); this shows
\begin{lemma}
$\|Y\|_X$ and $\|Y\|_0$ are equivalent norms.
\label{easy3}
\end{lemma}
For,
\begin{eqnarray*}
\|Y\|_X&=&\|R_X^{1/2}H_0^{1/2}R_0^{1/2}YR_0^{1/2}H_0^{1/2}R_X^{1/2}\|\\
&\leq&\|R_X^{12}H_0^{1/2}\|^2\|Y\|_0
\end{eqnarray*}
and the inequality in the other direction is similar. This equivalence
is the key to the extension of the manifold to other patches.

We see from the resulting estimate
\begin{equation}
Z_X-Z_{X+Y}={\rm Tr}\,e^{-H_X}-{\rm Tr}\,e^{-H_{X+Y}}\leq C\|Y\|_X
\end{equation}
that the partition function $Z_X$ is a Lipschitz function of
$X\in{\cal M}_0$.

\section{Affine Geometry in ${\cal M}_0$}
By an {\em affine structure} for a manifold ${\cal M}_0$ we mean a rule
for forming the convex mixture ``$\lambda\rho_1+\lambda^\prime\rho_2$''
($0\leq\lambda\leq 1;\rho_1,\rho_2\in{\cal M}_0$). An affine space is a
space with a specified affine structure; it is necessarily convex. The
unit ball $\widehat{\cal T}_1(0)$ in $\widehat{\cal T}(0)$ is a convex
subset of a Banach space and so has a natural affine structure coming from
the linear structure. By `transfer of structure', the chart $X\mapsto
\rho_X$ from $\widehat{\cal T}_1(0)$ to ${\cal M}_0$ provides ${\cal M}_0$
with the induced affine structure. This is called the canonical or
$(+1)$-affine structure. Clearly,
``$\lambda\rho_X+\lambda^\prime\rho_Y$'' $=\rho_{\lambda X+\lambda^
\prime Y}$
which differs from the usual mixture of states $\rho=\lambda\rho_X+
\lambda^\prime\rho_Y$. The latter is called the mixture or $(-1)$-affine
structure of the state space. While it is obvious that ${\cal M}_0$ is
$(+1)$-convex, it is not clear that it is $(-1)$-convex. That is,
while the $(-1)$-mixture $\rho$ of $\rho_1$ and $\rho_2$ is certainly a
state, it might not lie in ${\cal M}_0$ even if $\rho_1,\rho_2\in{\cal
M}_0$. 
\subsection{The $(+1)$-affine connection}
Let ${\cal T}_1$ and ${\cal T}_2$ be affine spaces; then an affine map
$U:{\cal T}_1\rightarrow{\cal T}_2$ is a map obeying
\begin{equation}
U\left(\lambda\rho_1+(1-\lambda)\rho_2\right)=\lambda U\rho_1+(1-\lambda)
U\rho_2
\end{equation}
for all $\rho_1,\rho_2\in{\cal T}_1$ and $0\leq\lambda\leq 1$.
An affine connection on a Banach manifold is an assignment, for each
continuous curve $L$ from $\rho_1$ to $\rho_2$, of an affine map
$U_L$ from the tangent space at $\rho_1$ to the tangent
space at $\rho_2$, obeying $U_LU_{L^\prime}=U_{L\cup L^\prime}$;
the map $U$ for the empty path $\emptyset$, when $\rho_1=\rho_2$, is
the identity,
and the symbol $L\cup L^\prime$ denotes the path $L$ followed by the path
$L^\prime$; if $L^\prime$ is the path $L$ with reversed parameter, we take
$L\cup L^\prime=\emptyset$. These axioms ensure that $U_L$ is an
invertible map for any $L$.
An affine connection is linear if $U_L(0)=0$ for every curve $L$; a linear
connection is torsion-free. If $U_L$ is independent of $L$ then the
connection is called {\em flat}, or curvature-free. The commonly used
formulation is the infinitesimal version of the above, obtained by
differentiating, if the structure is smooth.

In order to define the $(+1)$-affine connection concretely, we first put
coordinates on the tangent space at any $X\in{\cal M}_0$. 
We have seen that $\rho_0.Y$ is continuous in $Y\in{\cal M}_0$. Since
$H_X$ inherits all relevant properties of $H_0$, we obtain a similar
estimate $|\rho_X.Y|\leq{\rm const}\|Y\|_X$. The set
\begin{equation}
\widehat{\cal T}(X):=\{Y:\rho_X.Y=0\}
\end{equation}
is therefore a closed subspace of ${\cal T}(0)$ in the equivalent topology
defined by $\|\bullet\|_X$. We identify the tangent space at $\rho_X$
to be the Banach space $\widehat{\cal T}(X)$ with the norm $\|Y\|_X$.
We then take the $(+1)$-parallel transport $U_L$ of $Y-\rho_0.Y
\in\widehat{\cal T}
(0)$ along any path $L$ in the manifold to be the point $Y-\rho_X.Y\in
\widehat{\cal T}(X)$. This map takes $Y=0$ to zero, and extends to a linear
mapping from $\widehat{\cal T}(0)$ onto $\widehat{\cal T}(X)$. We see that
parallel transport is nothing other than the moving of the representative
point in the line $\tilde{\rho}$ from one hyperplane to the other.
Since this transport is independent of $L$ and linear, the connection
is flat and torsion-free.

\section{Extension of the Manifold}
We see two ways of extending the manifold by gluing new patches. The first
is to try to include as many $q_0$-small perturbations $X$ as
possible, and not just those
obeying $\|X\|_0<a_o$; recall that this condition is sufficient for
$X$ to have $q_0$-bound less than $a_o<1$. The second, and main,
method of extension, is to use any point $\rho_X$ in the first patch,
and to consider perturbations $Y$ of $H_X$ with $\|Y\|_X<a_{\rm x}=1-\beta_{_X}$.
This might lead out of ${\cal M}_0$; we can continue indefinitely,
starting at $H_Y$ etc. In this way we include eventually a state of any
temperature, and the manifold points generally in the $+H_0$
direction.

Suppose then that $X$ is symmetric and $q_0$-small enough. Then there exists
a self-adjoint operator $H_0$ whose form is $q_0+X$ with form-domain $Q_0$
and
\begin{equation}
|X(\psi,\psi)|\leq a\left(q_0(\psi,\psi)+b\|\psi\|^2/a\right)\mbox{ for
some }a<a_o.
\end{equation}
Let $\tilde{H}_0=H_0+\frac{b}{a}I$; this is self-adjoint on $D(H_0)$
and
\[|X(\psi,\psi)|\leq a\langle \tilde{H_0}^{1/2}\psi,\tilde{H_0}^{1/2}
\psi\rangle.\]
Let $\tilde{R}_0=\tilde{H_0}^{-1}$. Then for $\psi\in Q_0$,
we have for $\psi\in{\cal H}$,
\begin{equation}
\tilde{R}_0^{1/2}X\tilde{R}_0^{1/2}(\psi,\psi)\leq a\langle\tilde{R}^{1/2}
\psi,\tilde{H}_0\tilde{R}_0^{1/2}\psi\rangle=a\|\psi\|^2,
\end{equation}
so
\[\|X\|_{\tilde{0}}:=\left\|\tilde{R}_0^{1/2}X\tilde{R}_0^{1/2}\right\|
\leq a<a_o.\]
We have thus shown that for any $q_0$-small form $X$ with bound $<a_o$
there exists a choice of Hamiltonian $\tilde{H}_0$ such that $\rho_X$ lies
inside the open ball $\|X\|_{\tilde{0}}<a_o$.
Let us furnish $\widehat{\cal T}(0)$ with this
norm; it is equivalent to the norm $\|X\|_0$, since $\tilde{R}_0^{1/2}H_0
^{1/2}$ and $R_0^{1/2}\tilde{H}_0^{1/2}$ are both bounded. We can
therefore add the patches got in this way to the first patch, to get a
Banach manifold. We can enlarge the manifold even further, by analogy
with the classical case \cite{Pistone}. Let $X$ be a $q_0$-small form;
it therefore defines a unique pair of self-adjoint operators, $H_{\pm}$,
with forms $q_{\pm}:=q_0\pm X$; we include $\rho_X$ in the first patch
if for each choice of sign,
$\rho_{\pm}=\exp -H_{\pm}$ is of trace class.
For such an $X$ the quantum analogue of the Luxemburg norm is finite.
\begin{definition} We put
\[ \|X\|_L=\inf\left\{r>0:{\rm Tr}\left[\left(\exp-(H_0+X/r)+\exp-(H_0-X/r)\right)/
(2Z_0)\right]-1<1\right\}.\]
\end{definition}
For large $r$, $X/r$ is $q_0$-small, and so the traces make sense; since
$Z_X$ is continuous in $X/r$ if it is small, the set in the
infimum is non empty. As $r$ becomes smaller, either the operator $H_{\pm}$
fails to be unique, or the finiteness of the trace
might fail; in either case we put the trace equal to $\infty$, and the
corresponding $r$ is a lower bound for $\|X\|_L$.
It can be shown that $\|X\|_L$ is a seminorm.

The second extension of the manifold is to construct a similar chart
around a state $\rho_X$ as we did around $\rho_0$, where $X$ is
$\tilde{q}_0$-small, with bound $<a_o$. Since the choice of
$H_0$ was anyway arbitrary provided $H_0\geq I$, we drop the tilde; so
we consider $X\in{\cal M}_0$. Choose
$H_X\geq I$. This Hamiltonian inherits all the properties of $H_0$.
Let $Y$ be $q_X$-small with bound $<a_{\rm x}$; then there is a unique
self-adjoint operator $H_{X+Y}$, whose form domain is $Q_0$, and whose
form is $q_0+X+Y$, such that
\begin{equation}
\rho_{X+Y}=Z_{X+Y}^{-1}e^{-H_{X+Y}}
\label{XandY}
\end{equation}
is of trace-class, and $\rho_X.Y$ can be defined as
${\rm Tr}\left(\rho_X^\lambda Y\rho_X^{\lambda^\prime}\right)$.
Let ${\cal T}(X)$ be the Banach space of forms $Y$ such that $\|Y\|_X<
\infty$, with this norm. Since $\|\bullet\|_0$ and $\|\bullet\|_X$
are equivalent norms, this space is actually the same as ${\cal T}(0)$
as a set. The interior of this ball in ${\cal T}_X$ consists of certain
$q_X$-small forms which are $q_0$-bounded but might not lie in ${\cal
M}(0)$.
Let ${\cal M}_X$ be the set of states of the form eq.~(\ref{XandY}).
Again, two forms that differ by a multiple of $I$ yield the same state,
and there is a bijection between ${\cal M}_X$ and the set of lines
$\{\tilde{\rho}_{X+Y}\}$ in ${\cal T}_X$ parallel to $I$ that cut the
open ball in ${\cal T}(X)$ of radius $1-\beta_X$. The set of such lines is
furnished with the quotient topology. Let $\widehat{\cal T}(X)$ be the
(closed) hyperplane
$\{T\in{\cal T}(X):\rho_X.Y=0\}$. Each line $\tilde{\rho}_{X+Y}$ cuts
this hyperplane in a unique point, and those in the neighbourhood
of $Y=0$ cut the plane inside our open ball $\widehat{\cal T}_{a_{\rm x}}
(X)$. This gives a chart from an open set in ${\cal M}_X$ onto this 
ball. Again, the tangent space at $\rho_X$ is identified with
$\widehat{\cal T}(X)$. 
The $(+1)$-affine structure
in ${\cal M}_X$ is that induced from the linear structure of ${\cal T}(X)$.
We can enlarge this piece of the manifold to include all $q_X$-small
perturbations $Y$ such that $\exp-\{H_0+X+Y\}$ is of trace class,
and can cover the
enlarged set of states by consistent overlapping charts, in the same way
as for the first method of extension of ${\cal M}_0$.

The next step in building the manifold is to consider the union of ${\cal
M}_0$ and ${\cal M}_X$. The two charts are topologically compatible,
in that in the overlap ${\cal M}_0\cup{\cal M}_X$ the two norms $\|\bullet
\|_0$ and $\|\bullet\|_X$ induced by the charts are equivalent; see
lemma~(\ref{easy3}). The $(+1)$-affine structure of ${\cal M}_0$ and
${\cal M}_X$ are the same on their overlap, since both are induced by the
linear structure of ${\cal T}(0)$. Our choice of parallel transport
with the first patch reflects this, and can be extended in stages to
a transport between any two points in the union of the pieces.
Indeed, let $\widehat{X}_1$, $\widehat{X}_2$ lie in $\widehat{\cal T}(0)$,
so their means in $\rho_0$ are zero. Put $\widehat{Z}=\lambda\widehat{X}_1
+\lambda^\prime\widehat{X}_2$, and let $U$ denote the parallel transport
from $\rho_0$ to $\rho_X$. Then
\[U\widehat{X}_i=\widehat{X}_i-\rho_X.\widehat{X}_i,\;\;i=1,2,\mbox{ and }
U\widehat{Z}=\widehat{Z}-\rho_X.Z\]
Then
\begin{eqnarray*}
U\left(\lambda\widehat{X}_1+\lambda^\prime \widehat{X}_2\right)&=&
\widehat{Z}-\rho_X.\widehat{Z}\\
&=&\lambda\widehat{X}_1+\lambda^\prime\widehat{X}_2-\rho_X.\left(\lambda
\widehat{X}_1+\lambda^\prime\widehat{X}_2\right)\\
&=&\lambda U\widehat{X}_1+\lambda^\prime U\widehat{X}_2.
\end{eqnarray*}
That is, $U$ takes the convex mixture in ${\cal T}(0)$ to that in ${\cal
T}(X)$. Thus the union of the first two pieces is a Banach manifold
furnished with a flat torsion-free affine structure and the $(+1)$-parallel
transport $U$.

We can extend further, to a third piece, starting from a different point
$X^\prime$ in ${\cal M}_0$ or from a point in ${\cal M}_X$ outside
${\cal M}_0$. In either case we arrive at a $q_0$-bounded form with
domain $Q_0$, and a third piece of the manifold with a chart into
an open ball of the Banach space $\{Y:\rho_{X^\prime}.Y=0\}$,
with norm $\|\bullet\|_{X^\prime}$ equivalent to the norms already defined.
We continue by induction, starting at any point in the manifold
obtained already, to get to any $q_0$-bounded form that can be arrived at
in a finite number of steps. At each stage, starting from $\rho_X$
we enlarge the ball of radius $a_{\rm x}$ by the first method, to include
all $q_X$-small forms which define a state. 
Moreover, suppose we arrive at two far points, $H_0+X$ and $H_0+Y$, which
however lie in each other's patch. When neither
$X$ nor $Y$ is $q_0$-small (but are $q_0$-bounded), we can by
construction find a finite chain $X_1,X_2\ldots$
from $X$ to $H_0$ and 
another finite chain $Y_1,Y_2\ldots$ from $H_0$ to $Y$, each
small relative to the last; then $R_X^{1/2}YR_X^{1/2}$
is a finite product
\[R_X^{1/2}H_{X_1}^{1/2}\;R_{X_1}^{1/2}H_{X_1}^{1/2}\ldots H_0^{1/2}R_{Y_1}^
{1/2}\;H_{Y_1}^{1/2}R_{Y_2}^{1/2}\ldots R_{Y_n}^{1/2}H_YR_{Y_n}^{1/2}
\ldots R_X^{1/2}\]
which is bounded. Thus $\|Y\|_X$ is finite. Similarly $\|X\|_Y$ is finite.
Thus when we then extend to all states obtainable in this way in a finite
number of steps, all the norms of any overlapping region are equivalent.
With each enlargement, we define the $(+1)$-affine structure
and parallel transport in stages from chart to chart, to give a flat
torsion-free connection.
\begin{definition}
The information manifold ${\cal M}$ defined by $H_0$ consists of all states
obtainable in a finite number of steps, by extending from ${\cal M}_0$ by
either the first method or the second method, as explained above.
\label{manifold}
\end{definition}
The Cramer class of each $\rho\in{\cal M}$ is the set of $q_0$-bounded
forms.

The question now arises, when we add perturbations $X_1,\ldots,X_n$
and $Y_1,\ldots,Y_m$ as above, and $X_1+\ldots+X_n=Y_1+\ldots+Y_m$ as
forms (on $Q_0$), whether we arrive at the same state whichever route we
take. We do, since there is a unique self-adjoint operator defined by the
form
\[q_0+X_1+\ldots+X_n=q_0+Y_1+\ldots+Y_m\]
with form domain $Q_0$.

We now have a natural result.
\begin{theorem}
${\cal M}$ is $(+1)$-convex.
\end{theorem}
PROOF.
We first prove the result when $\beta_0=0$.
Then the only condition on the size of a perturbation $Y$ of $H_X$
is that it be $q_{_X}$-small. In this case it is obvious that the manifold is
a cone.

Let ${\cal M}(0)$ denote the set of states $\rho_X$ where $X$ is
$q_0$-small. Then we define ${\cal M}(n)$ inductively by
\begin{definition}
$\rho_X\in{\cal M}(n)$ if there exists $\rho_Y\in{\cal M}(n-1)$ such that
$X-Y$is $q_Y$-small, where $q_Y=q_0+Y_1+\ldots+Y_{n-1}$,
and each addition $Y_j$ is small relative to $q_{j-1}$.
\end{definition}
We show that ${\cal M}(0)$ is $(+1)$-convex, and that if ${\cal M}(n-1)$
is $(+1)$-convex, so is ${\cal M}$.

Suppose then that $X_i\in{\cal M}_0$, $1=1,2$. Then for $\psi\in Q_0$,
\begin{eqnarray*}
|X_1(\psi,\psi)|&\leq&a_1q_0(\psi,\psi)+b_1\|\psi\|^2\\
|X_2(\psi,\psi)|&\leq&a_2q_0(\psi,\psi)+b_2\|\psi\|^2.
\end{eqnarray*}
Then
\begin{eqnarray*}
\left|\left(\lambda X_1+\lambda^\prime X_2\right)(\psi,\psi)\right|&\leq&
\lambda\left|X_1(\psi,\psi)\right|+\lambda^\prime\left|X_2(\psi,\psi)\right|
\\
&\leq&\lambda\left(a_1q_0(\psi,\psi)+b_1\|\psi\|^2\right)+\lambda^\prime
\left(a_2q_0(\psi,\psi)+b_2\|\psi\|^2\right)\\
&\leq& aq_0(\psi,\psi)+b\|\psi\|^2
\end{eqnarray*}
where $a=\max\{a_1,a_2\}<1$ and $b=\max\{b_1,b_2\}$. Hence $\lambda X_1
+\lambda^\prime X_2$ is $q_0$-small, and ${\cal M}(0)$ is convex.

Now let ${\cal M}(n)$ be obtained from ${\cal M}(n-1)$ as in the definition
~(\ref{manifold}). So let $q_Y$ be of the form $q_0+Y$ where $\rho_Y\in
{\cal M}(n-1)$, and let $X$ be $q_Y$-small. Then $\rho_{X+Y}\in{\cal M}(n)$,
and any element of ${\cal M}(n)$ is of this form. Let $\rho_1,\rho_2\in
{\cal M}(n)$; then there exist $Y_1,Y_2$ such that $\rho_{Y_1},\rho_{Y_2}
\in{\cal M}(n-1)$, and writing $q_1=q_0+Y_1$ and $q_2=q_0+Y_2$, then
there exist forms $X_1,X_2$ such that $X_1$ is $q_1$-small and $X_2$ is
$q_2$-small, and $\rho_1$, $\rho_2$ are the states corresponding to
$q_1+X_1$ and $q_2+X_2$. Let
$q=\lambda q_1+\lambda^\prime q_2$; the state corresponding to $q$ lies in
${\cal M}(n-1)$, since this is $(+1)$-convex, by the inductive hypothesis.
A simple estimate shows that $\lambda X_1+\lambda^\prime X_2$ is
$q$-small,
so that the state corresponding to $q+\lambda X_1+\lambda^\prime X_2$
lies in ${\cal M}(n)$. But the latter is $\lambda(q_1+X_1)+\lambda^\prime
(q_2+X_2)$, whose corresponding state is the $(+1)$ mixture of $\rho_1$ and
$\rho_2$. This shows that ${\cal M}(n)$ is
$(+1)$-convex.

Now relax the condition that $\beta_0=0$, and define the part ${\cal M}(n)$
to be the set of states obtained from ${\cal M}(n-1)$ as above, but using
only sufficiently small perturbations. Then all the conclusions derived
above remain true, up to the result
that $\lambda X_1+\lambda^\prime X_2$ is $q$-small. Thus $q+\lambda X_1+
\lambda^\prime X_2$ is the form of a self-adjoint operator that is bounded
below; call this operator $H$. Now, by the convexity of $Z_X$, $\exp -H$
is of trace class, since $\rho_1$ and $\rho_2$ are. The same is true if we
replace $X_i$ by $-X_i$; hence $Z^{-1}\exp -X$ lies in ${\cal M}(n)$, by
the first method of extension. \hspace{\fill}$\Box$

We have not been able to prove that the manifold is $(-1)$-convex; if
$\rho_1$ and $\rho_2$ are density operators in the first patch,
then obviously, $\rho:=\lambda\rho_1+\lambda^\prime\rho_2$ is a density
operator. All we can show
from the operator convexity of $-\log x$ \cite{Bendat,Lindblad}, is that
$-\log\rho=H_0+X$, where $X$ has $q_0$-bound 1; but we need the bound
to be {\em less than} 1 for $\rho$ to lie inside the first patch.
\section{\bf Outlook}
We have defined a Banach manifold, with the flat torsion-free
$(+1)$-connection. The canonical variables at $\rho_0$, are the
centred $q_0$-bounded forms $X$, with the norm $\|\bullet\|_0$.
These are $(+1)$-affine coordinates, and the manifold is a convex set
when expressed in terms of these. 
The Massieu function $\Psi$ is a continuous convex
function on the manifold. We can therefore construct the Legendre
transform using Fenchel's theory, to obtain the `mixture' variables
$\rho_Y.X$ at any point $\rho_Y$ in the manifold. The entropy is a
continuous function. With more regularity, we have been able to
show that the {\em BKM} metric is finite at regular points, and
is the Fr\'{e}chet derivative of $\rho_Y.X$, as in the classical and
finite-dimensional cases. Moreover, the free energy is real-analytic.
This work, \cite{RFSMRG,RFS4} which extends \cite{RFS1},
will be published elsewhere.

\hspace*{\fill}{\bf Acknowledgements}\hspace{\fill}$\left.\right.$

\noindent This work is part of an ongoing programme being worked out with
P. Combe, G. Burdet and H. Nencka, of CPT and CNRS Marseille and the
University of Madeira. I thank P. Combe for arranging the visit to
Marseille, and CNRS for financial support; I thank H. Nencka for arranging
the visit to Madeira. I am endebted to E. B. Davies for help with the proof
of Lemma 4.

%\newpage
%\vspace*{5in}
%\hspace*{\fill}Two Overlapping Patches\hspace{\fill}$\left.\right.$
\end{document}